# Dynamic nanoscale structural correlations in strontium ruthenate

M. Spaić[1], R. Spieker[2], I. Bilonić[1], A. Paul[3,4], B. Krohnke Orquera[2], X. He[2], E. Topić[5], A. Minelli[6], F. Ye[6], N. Kikugawa[7], D. A. Sokolov[7], M. J. Krogstad[8], S. Rosenkranz[9], R. Osborn[9], T. Birol[4], M. Greven[2*], and D. Pelc[1,2*]

[1]Department of Physics, Faculty of Science, University of Zagreb, HR-10000 Zagreb, Croatia

[2]School of Physics and Astronomy, University of Minnesota, Minneapolis, MN 55455, USA

[3]Department of Physics, Indian Institute of Technology Jammu, Jammu and Kashmir 181221, India

[4]Department of Chemical Engineering and Materials Science, University of Minnesota, Minneapolis, MN 55455, USA

[5]Department of Chemistry, Faculty of Science, University of Zagreb, HR-10000 Zagreb, Croatia

[6]Neutron Scattering Division, Oak Ridge National Laboratory, Oak Ridge, TN, USA

[7]Max-Planck-Institute for the Chemical Physics of Solids, Nöthnitzer Str. 40, 01187 Dresden, Germany

[8]Advanced Photon Source, Argonne National Laboratory, Lemont, IL, 60439, USA

[9]Materials Science Division, Argonne National Laboratory, Lemont, IL, 60439, USA

*correspondence to: greven@umn.edu, dpelc@phy.hr

**Strontium ruthenate ($Sr_2RuO_4$, SRO) has been the subject of intense research as a model quasi-two-dimensional metal with strong electronic correlations and potential exotic multi-component superconductor. Yet the nature of the superconducting state and its emergence remain debated, despite highly detailed knowledge of the normal-state electronic properties. Here we use diffuse neutron and x-ray scattering to uncover self-organized structural fluctuations on the picosecond timescale in SRO. We show that these nanoscale correlations do not originate from extrinsic disorder but rather involve cooperative displacements of oxygen atoms in the quintessential $RuO_2$ planes. Moreover, the observed displacement pattern is consistent with distortions due to incipient orbital order that we obtain in first-principles calculations, which suggests that orbital effects could play a pivotal role in the physics of SRO. Similar dynamic correlations may play a role in the physical properties of a wide range of prominent oxides with closely-related lamellar structures, such as the cuprates and nickelates.**

During the past three decades, SRO has been heavily studied as a candidate multi-component superconductor with spin-triplet pairing [1-4]. The superconducting state in SRO is highly sensitive to lattice defects [5], and the transition temperature can be significantly enhanced via uniaxial strain [6-9], which has spurred efforts to obtain single crystals with extremely low concentrations of lattice defects. Yet studies of superconducting fluctuations indirectly point to the presence of intrinsic inhomogeneity, as revealed through an unusual, extended fluctuation regime [10,11]. In addition, charge transport [12] and optical [13] measurements in the metallic normal state have shown evidence for spontaneous orientational symmetry breaking in thin-film samples, and it remains unknown if such effects are present in bulk crystals. It is also known that a structural distortion occurs at the free surface of SRO [14], and recent *ab initio* calculations have uncovered a tendency toward orbital order, which would result in Jahn-Teller-like distortions of the Ru-O plaquettes [15]. Given the stoichiometric nature of SRO and the absence of bulk structural instabilities, SRO is an excellent model system for the study of intrinsic nanoscale structural correlations and their interplay with the electronic subsystem.

Here we leverage recent developments in state-of-the-art diffuse neutron and x-ray scattering methodology [16] to investigate correlated short-range deviations from the average crystal structure in SRO. When atoms are displaced from their expected high-symmetry positions in the crystalline lattice, diffuse scattering appears away from the Bragg peaks, and the atomic displacement patterns can be reconstructed from the scattering data. We uncover structured diffuse scattering in SRO, which shows that self-organized inhomogeneity can form even in a pristine system such as SRO. Furthermore, we perform *ab initio* calculations to confirm the tendency towards orbital order and determine the associated lattice distortions, and we find that the atomic displacements are similar to the displacement pattern that we obtain from diffuse scattering data analysis. This result implies that the system is close to an orbital instability, with important implication for electronic physics.

It has long been known that the tetragonal average structure of SRO (Fig. 1a) is stable down to the lowest temperatures [17], which enables straightforward reconstruction of atomic displacements from the diffuse scattering data. We combine neutron and x-ray scattering to take advantage of the different cross-sections of the two probes for different atomic species; most importantly, the relative scattering strength of oxygen is much higher for neutrons than for x-rays, which makes neutron scattering a better probe of oxygen displacements [18]. Indeed, our experiments reveal the presence of correlations that only involve the oxygen atoms: no discernible diffuse signals are observed in the x-ray data, whereas the neutron measurements clearly show diffuse scattering at non-integer reciprocal space positions (Fig. 1b,c). Two types of features are seen: diffuse peaks close to (½ ½) (half-integer) positions of the tetragonal (I4/mmm) reciprocal lattice, and diffuse clouds around some of the regular Bragg peaks. Closer inspection of the half-integer diffuse intensities reveals the absence of any signal at several such positions, most notably the reflections with equal in-plane wavevectors (green circle in Fig. 1b). In addition, the diffuse signals show strong asymmetry: they almost exclusively appear on the high-wavevector side of the integer and half-integer positions (inset in Fig. 1b, and Fig. 1d). Finally, the distortions are only weakly correlated between $RuO_2$ planes, since the dependence of the diffuse intensity on the out-of-plane wavevector does not show sharp peaks along the out-of-plane direction $L$ (Fig. 1e).

These strong constraints allow us to determine the nature of the real-space atomic displacements directly from the data. Of the four irreducible distortions of the oxygen plaquette that double the unit cell (Fig. 2a), it can be demonstrated through a simple structure-factor calculation (see also Methods) that only two show extinctions at diagonal half-integer positions: the rotational distortion M1 and the rhombic distortion M4. The total displacement must then be a sum of M1 and M4. In order to determine the ratio of their weights, we perform a structure factor refinement on nine distinct half-integer peaks (see Methods for details) which shows that the most likely distortion is an equal-weight superposition, i.e., a shear-like distortion (Fig. 2b). In turn, the asymmetry of diffuse features around both integer and half-integer positions implies that the atomic displacements must include a local decrease of oxygen-oxygen distances. This can only occur if the correlated displacements include an extended defect, i.e., an antiphase boundary, at which the shear-like pattern reverses and the oxygen sublattice shows an overall contraction (Fig. 2d). We thus deduce that the diffuse scattering originates from regions with correlated shear-like distortions and an antiphase boundary that are embedded in the tetragonal lattice of SRO. Given that the width of the reciprocal-space features is around 0.1 r.l.u. (Fig. 1d) or 0.16 $Å^{-1}$, the lateral spatial extent of these regions is on the order of ten unit cells, and possibly significantly more, since it is difficult to reliably determine the detailed shape of the diffuse signals.

The CORELLI diffuse neutron scattering spectrometer [19] permits a qualitative determination of the dynamic properties of the structural correlations. The scattering is separated into two channels: quasi-elastic (i.e., neutron energy transfer below 0.3-2 meV; the resolution decreases with increasing wavevector/neutron incident energy) and "total" scattering (energy-integrated up to ~10 meV). Interestingly, diffuse features in the quasi-elastic channel are barely visible even at the largest wavevectors (Extended Data Fig. 3), whereas the signal is much stronger in the energy-integrated channel (Fig. 1d), implying that the correlations are dynamic. Since the features become detectable in the quasi-elastic channel in a wavevector region where the energy resolution is around 2 meV, the energy scale of the nanoscale fluctuations is a few meV, and the corresponding timescale therefore is in the picosecond range. In order to quantitatively compare the observed features to phonon scattering, we compute the phonon contribution, known as thermal diffuse scattering (TDS). For the phonon dispersions we use a first-principles calculation of the phonon spectrum of SRO (see Methods and Fig. 3a) that is in excellent agreement with detailed high-resolution inelastic neutron scattering measurements [17,24]. The softest zone-boundary (i.e., unit-cell-doubling) phonon at around 8 meV corresponds to the M1 mode [17], and this phonon shares some similarities with the displacements that we observe. Yet there are also crucial differences in the diffuse scattering: the above-mentioned reciprocal space asymmetry is not present in the calculated TDS, nor is any asymmetry observed in the phonon dispersions [17,24]; the phonon scattering yields higher intensities at forbidden integer Bragg positions; and the phonon contribution shows a decrease with cooling due to the Bose-Einstein occupation factor. In contrast, once the symmetric phonon part is subtracted, the remaining diffuse intensity shows an increase with cooling (Fig. 3b). From diffuse scattering alone it is difficult to determine if the increase originates from changes in the sample volume fraction where the inhomogeneity is present, an increase of the atomic displacement amplitudes, a softening of the local excitations, or a combination of these effects; dedicated energy-resolved measurements are needed to elucidate the evolution of the short-range correlations with energy and temperature.

We now discuss possible origins of the diffuse signals. Although the neutron scattering features could in principle originate from low-energy magnetic excitations that are known to be present in SRO [20,21], this possibility is excluded by the fact that the diffuse intensities increase with increasing wavevector, while the opposite behaviour is expected for magnetic scattering; within experimental sensitivity, we do not see magnetic signals at low wavevectors. Moreover, since the superconducting transition temperature of SRO is very sensitive to point defects, and our samples have $T_c$ values close to the maximum achievable value of 1.5 K, the crystals have an extremely low point defect concentration, and such defects cannot be the cause of the observed features. Yet extended defects such as dislocations may play a role: dislocations are associated with long-range strain fields [22] that could induce local softening of the octahedral rotation mode and lead to an enhanced diffuse scattering signal. In order to quantify this possibility, we perform high-resolution x-ray diffraction measurements on several SRO crystals, from the batch used for the neutron scattering measurement, and combine them with *ab initio* calculations of the phonon spectrum with strain. The calculations show significant effects on the softest M1 mode for tensile *c*-axis strains of order 1% (Fig. 2d), but the diffraction profiles suggest that the volume fraction of the material that is subjected to such strains is negligible (Fig. 2e). Importantly, this is consistent with magnetic susceptibility measurements of superconducting fluctuations, where signals that have been interpreted as originating from rare regions with nonzero strain are observed significantly above the bulk $T_c$ [10,11]. Those signals are also at the sub-1% level, which suggests that some internal strain is present, but likely not responsible for the diffuse neutron scattering: although the scattering from soft phonons in the strained regions would be enhanced in inverse proportion to their energy, this is not enough to compensate for the small volume fraction. Such phonon-related scattering would also not explain the reciprocal space asymmetry.

It is a distinct possibility that the nanoscale distortions are related to electronic orientational symmetry breaking found in thin films of SRO [12,13]. Notably, the magnitude of those features decreases with increasing temperature, in agreement with our observations. While such electronic symmetry breaking would not necessarily lead to observable atomic displacements in the bulk, fluctuating domain boundaries would be consistent with the proposed defect structure, with local electrostatic interactions leading to the displacement pattern in Fig. 2d. However, the symmetry-breaking direction would need to be along the in-plane diagonal to yield the monoclinic-type distortion through electrostatic interactions, while optical reflectivity experiments on thin films [13] and existing mean-field calculations based on the three-band Hubbard model [13,23] suggest alignment with the Ru-O bonds.

Detailed measurements of the SRO phonon characteristics suggest that electron-phonon coupling to several modes, including the rotational mode, is strong [17,24]. In addition, the electronic structure of SRO is close to a Lifshitz transition, with pronounced Fermi-surface nesting and a large density of states, which is also known to affect the static elastic properties [25]. As noted, an orbitally ordered state with Jahn-Teller distortions in the Ru-O planes was recently uncovered [15], and we reproduce the orbital order in our *ab initio* calculations (Fig. 2c). Moreover, we obtain an *ab initio* estimate of the oxygen displacement amplitude of ~0.01 lattice units, in excellent agreement with the value extracted from the structural refinement of diffuse scattering (see Methods). We thus believe that it is most likely that the observed inhomogeneity originates from interactions between the lattice and the electronic subsystem. Along with the close correspondence between the predicted Jahn-Teller distortion and the experimentally deduced displacement pattern, the low energy and small reciprocal space width of the diffuse peaks are consistent with slow, cooperative orbital-lattice fluctuations that involve several unit cells. The proposed structure (Fig. 2d) would then naturally correspond to a nanoscale orbitally ordered region with an antiphase boundary. Note also that the *ab initio* calculations only show orbital order in the presence of magnetic order (either ferro- or antiferromagnetic) of Ru spins, and a targeted search for weak diffuse magnetic scattering, e.g., with polarized neutrons, could therefore yield further insight.

The finding of intrinsic dynamic inhomogeneity has far-reaching implications for the physics of SRO and other complex oxides. First, it shows that the crystalline lattice seen locally by some of the electronic charge carriers is significantly different from the average structure: the symmetry is lowered, and a lattice contraction appears. Both effects are important for SRO superconductivity, as it has been suggested that spatially inhomogeneous strain can fundamentally modify the local superconducting pairing symmetry [26]. In particular, the presence of fluctuating orbital order and associated antiphase boundaries could explain some of the major mysteries of SRO physics and provide an important ingredient to understand the superconducting mechanism. In perovskites with antiferromagnetic-like spin configurations, zone-boundary distortions that reduce the translational symmetry so that the magnetic unit cell matches the crystallographic unit cell can lead to altermagnetism [27-29], and it is known that long-range magnetic order can be induced with modest strain in SRO [30]. A surface altermagnetic state has recently been proposed [31] to explain both polar Kerr effect [32] and muon spin rotation [30,33] measurements of weak time-reversal symmetry breaking below $T_c$, and our results suggest that such effects could also appear in the vicinity of antiphase boundaries in the bulk. It will be particularly interesting to investigate if external symmetry-breaking fields such as uniaxial strain can induce long-range orbital order and altermagnetism in SRO. Incipient orbital order would imply that electron-phonon coupling is strong and could play a role in superconductivity [15]; SRO has also been proposed as an example of a Hund's metal, with inter-orbital interactions as a key ingredient [34,35] and possible inter-orbital superconducting pairing [36,37]. Our results provide crucial experimental input to refine models and understand the role of orbital physics in SRO and other oxides with strong electronic correlations. More broadly, the presence of self-organized dynamical inhomogeneity in this pristine and structurally simple oxide indicates that such effects might be commonplace in similar materials, such as the isostructural cuprates and nickelates, but are likely masked by compound-specific idiosyncrasies such as substitutional disorder and structural phase transitions.

## Methods

*Samples.* Single crystals of SRO were grown using the traveling floating zone method similar to previous work. Samples for diffuse neutron and x-ray scattering had masses of 200 mg and 2 mg, respectively. High-resolution x-ray diffraction measurements were performed at ambient temperature on the (0 0 14) Bragg peak using a Panalytical Empyrean diffractometer with high-resolution optics; the (440) Bragg peak of a germanium single crystal was used as a reference, with Bragg peak widths around $10^{-3}$ Å$^{-1}$ (see Extended Data Fig. 4).

*Diffuse neutron and x-ray scattering.* Transmission x-ray scattering experiments were performed at beamline 6-ID-D of the Advanced Photon Source, Argonne National Laboratory, using a photon energy of 87 keV and a CdTe area detector. Reciprocal space maps were reconstructed using NXRefine [38] from three sets of full rotation scans, with the rotation axis tilted by -15°, 0° and 15°, respectively, in order to improve statistics and eliminate detector gaps; for details of the reconstruction method see [39]. Each set of three scans is around 30 minutes. The base temperature of 30 K was achieved using a helium cryostream.

Neutron scattering measurements were performed using the CORELLI instrument [19] at the Spallation Neutron Source, Oak Ridge National Laboratory, with a cross-correlation chopper employed to separate quasielastic from energy-integrated scattering. The Mantid software package was used for data reduction, including Lorentz and spectral corrections [40]. To obtain full reciprocal space maps, the sample was rotated 360° along a vertical axis that approximately coincided with the (1 1 0) crystallographic direction, with 8-10 hour counting times for the entire rotation.

*Neutron data analysis.* All neutron diffraction analyses were performed using multidimensional histogram data stored in the NeXus format. A fundamental step across all datasets involved binning of reciprocal lattice coordinates (H, K, L) to ensure accurate spatial mapping. For every integration and binning step, statistical uncertainties were rigorously maintained by propagating values from the squared errors data. Final intensities are reported in arbitrary units, typically scaled for visual clarity, with error bars representing the standard deviation.

A central component of the analysis for all diffuse signals was the transformation of the reciprocal lattice coordinates (H, K) into a rotated frame ($x_{rot}$, $y_{rot}$) defined as:

$$x_{rot} = \frac{H+K}{\sqrt{2}} \qquad y_{rot} = \frac{H-K}{\sqrt{2}}$$

This coordinate transformation was used across all diffuse scattering analyses to extract one-dimensional cuts along the diagonal between two Bragg peaks.

For the comparison between total and quasielastic scattering (Fig.1d), the data obtained at 5 K were integrated over $L$ axis from -12.0 to 12.0 r.l.u.. For the quasielastic component, intensities were normalized by averaging the signal over all contributing voxels within each $y_{rot}$ bin. For the energy-integrated intensity, the minimum value was subtracted to account for background, and the value at the same wavevector was then subtracted from the quasi-elastic intensity. The variation of intensity with $L$ (Fig. 1e) was extracted by selecting a (H, K) square centered at (-1.5, 6.6) with a width of $\pm$ 0.05 r.l.u.. The data was binned along $L$ using a width of $\Delta$L = 0.2 r.l.u.. A spatial background profile was extracted by shifting the integration region to a nearby reciprocal space position at (-1.5, 5.9) for direct comparison.

A multicomponent fitting pipeline was implemented to compare experimental scattering at 5 K and 120 K with theoretical thermal diffuse scattering (TDS) profiles (Fig. 3a). The experimental signal was first heuristically modeled using a double Gaussian function combined with a 3rd-order polynomial,

$$I(q) = A_1 exp\left(-\frac{(q-q_{01})^2}{2\sigma_1^2}\right) + A_2 exp\left(-\frac{(q-q_{02})^2}{2\sigma_2^2}\right) + p_0 + p_1 q + p_2 q^2 + p_3 q^3$$

where $q$ is the relative wavenumber, and $A_i$, $q_{0i}$ and $p_{0i}$ are free parameters. The double Gaussian accounts for local scattering features, while the 3rd-order polynomial represents the background baseline and was subtracted from the raw data. The experimental profiles were then symmetrized around the (½ ½) point to isolate the symmetric phonon contribution, and the theoretical TDS profiles were scaled to the background-subtracted symmetrised data: $I_{sym} = A \cdot I_{TDS} + offset$, where the scaling factor $A$ is temperature-independent, and was determined from the 120 K dataset, while the offset is different for each temperature and obtained using least-squares minimization.

To obtain the temperature-dependent integrated intensity (Fig.3b), signals were symmetrized around the peak center and background-corrected by subtracting the minimum observed intensity. The symmetrized signal was then subtracted from the total, and the resulting intensity obtained by integration of points around the peak. A univariate spline was fitted to the experimental points to serve as a visual guide for the temperature trend. The phonon line corresponds to a Bose-Einstein distribution factor $\frac{1}{1-\exp\left(\frac{-\hbar\omega}{T}\right)}$ with phonon energy 8 meV.

*Structure factor calculation.* For convenience, we use an orthorhombic unit cell of the SRO structure (space group Bmab) for the structure-factor calculation; this does not affect the end results, and several other choices of unit cell are equivalent. In Extended Data Fig. 2 we transform the Bragg peak positions back into the tetragonal structure (I4/mmm). The oxygen displacements corresponding to irreducible distortions M1 and M4 are

$$\delta_{O1} = \begin{pmatrix} M_1 + M_4 \\ M_1 - M_4 \end{pmatrix}, \delta_{O2} = \begin{pmatrix} -M_1 - M_4 \\ M_1 - M_4 \end{pmatrix}, \delta_{O3} = \begin{pmatrix} -M_1 - M_4 \\ -M_1 + M_4 \end{pmatrix}, \delta_{O4} = \begin{pmatrix} M_1 + M_4 \\ -M_1 + M_4 \end{pmatrix}$$

where $M_1$ and $M_4$ are the respective weights. If we define $\delta_+ = \delta(M_1 + M_4)/\sqrt{M_1^2 + M_4^2}$ and $\delta_- = \delta(M_1 - M_4)/\sqrt{M_1^2 + M_4^2}$, with $\delta$ an overall displacement (in l.u.), the neutron structure factor can be written as

$$\begin{aligned} F_{HKL} = {} & 2b_O \cos 2\pi z_O L \left(1 + e^{i\pi(H+K)} + e^{i\pi(H+L)} + e^{i\pi(K+L)}\right) \\ & + 4b_O \cos\left(\frac{1}{2}\pi H + 2\pi K\delta_-\right)\cos\left(\frac{1}{2}\pi K - 2\pi H\delta_+\right) + 4b_0 e^{i\pi L}\cos\frac{1}{2}\pi H \cos\frac{1}{2}\pi K \\ & + b_{Ru}\left(1 + e^{i\pi(H+K)} + e^{i\pi(H+L)} + e^{i\pi(K+L)}\right) \\ & + 2b_{Sr}\cos 2\pi z_{Sr} L \left(1 + e^{i\pi(H+K)} + e^{i\pi(H+L)} + e^{i\pi(K+L)}\right) \end{aligned}$$

where $b_O$, $b_{Ru}$ and $b_{Sr}$ are the oxygen, ruthenium and strontium scattering lengths, respectively, and $z_O$ and $z_{Sr}$ are the apical oxygen and strontium positions along the crystallographic $c$-axis; we also assume that the distortions are uncorrelated between neighbouring Ru-O planes, which is approximately justified by the observed smooth $L$ dependence of the scattering; including a finite (positive or negative) correlation between neighbouring planes does not substantially alter the results. The fact that the intensity shows a broad peak at $L \sim 5$ r.l.u. suggests that apical oxygen atoms might shift as well, which is not surprising if the Ru orbital configuration changes, but we neglect this effect in the refinement to minimize the number of free parameters. In order to determine the weights $M_1$ and $M_4$, we perform a least-square fit of $|F_{HKL}|^2$ to the integrated intensities of nine half-integer peaks (see Extended Data Figure 2), with two additional free parameters: an overall scaling factor, and the displacement $\delta$. The minimization yields $M_1 = M_4 = 1$ and $\delta = 0.015(1)$ l.u., which corresponds to the shear-like distortion in Fig. 2a and suggests that the atomic displacements are on the order of 1% of the unit cell. Note that a similar structure factor calculation can be performed in a straightforward way for modes M2 and M3 to show that these displacements do not lead to extinctions at (H/2 H/2) type positions.

*High-resolution x-ray profiles.* The X-ray diffraction (XRD) analysis utilized a multi-step protocol to isolate data for different Cu emission lines and remove instrumental broadening. A critical step in the XRD analysis was the separation of diffraction from the $K\alpha_1$ and $K\alpha_2$ emission lines. The raw $2\theta$ scattering data was converted into reciprocal lattice units (Q) using the characteristic wavelengths for each component respectively: $\lambda_1 = 1.540562$ Å, $\lambda_2 = 1.544398$ Å. Individual peak regions were extracted for each emission line and modeled using a Lorentzian function to determine precise peak centers, widths (HWHM), and amplitudes. While both components were analyzed to characterize the sample's structural properties, the $K\alpha_1$ component was selected for all final visualizations and subsequent deconvolution to provide the clearest representation of the diffraction features. The instrumental resolution function was determined from the measured diffraction profile of a high-quality germanium (Ge) reference. The isolated $K\alpha_1$ data for both the SRO samples and the Ge reference were then interpolated onto a uniform, linearly spaced grid to satisfy the requirements of the deconvolution algorithm. Richardson-Lucy deconvolution was applied iteratively (for 100 iterations) to mathematically extract the true sample signal from the measured resolution-broadened profile. The resulting deconvolution-corrected intensities were normalized to a 0–1 range and plotted on a logarithmic scale against $\Delta q$ (scaled by $10^3$). This allowed for the simultaneous observation of the high-intensity Bragg peak and the subtle diffuse scattering features in the tails down to intensities of $10^{-4}$.

*Ab initio calculations.* First-principles electronic structure calculations were carried out within the framework of density functional theory (DFT) using the VASP package [41,42]. The exchange-correlation energy was treated within the generalized gradient approximation using the PBEsol parametrization of the exchange-correlation functional [43]. Projector augmented wave (PAW) potentials [44] were used with valence configurations of $4d^7 5s^1$, $2s^2 2p^4$, and $4s^2 4p^6 5s^2$ for Ru, O, and Sr atoms, respectively. An energy cutoff of 560 eV was used to truncate the plane-wave basis set. Brillouin zone integrations were performed using a uniform mesh of $12 \times 12 \times 12$ k-points for structural optimization. Strain-dependent phonon frequencies of nonmagnetic $Sr_2RuO_4$ were calculated using the finite-displacement method as implemented in the phonopy package [45], employing a $2 \times 2 \times 2$ supercell constructed from the seven-atom rhombohedral unit cell. A C-type antiferromagnetically ordered state with an on-site Hubbard parameter $U = 2$ eV was employed to realize the antiferro-orbital ordering. A finite U value is necessary for the emergence of orbital ordering in magnetic phase.

*Thermal diffuse scattering*. To obtain the full phonon dispersions shown in Fig. 3a, electronic structure was calculated for a conventional tetragonal cell of $Sr_2RuO_4$ within density functional theory using the Quantum ESPRESSO software package [46], with the exchange-correlation energy described within the generalized-gradient approximation using the PBE functional [47]. Fully relativistic norm-conserving ONCVPSP pseudopotentials [48,49] were employed for all atomic species. In addition, spin-orbit coupling was included using a noncollinear formalism. The plane-wave kinetic-energy and charge-density cutoffs were set to 1361 eV and 6803 eV, respectively, and the Brillouin zone was sampled using a $10 \times 10 \times 3$ k-point mesh. The lattice dynamics of $Sr_2RuO_4$ was then calculated using density-functional perturbation theory as implemented in the PHonon module of Quantum ESPRESSO where dynamical matrices were evaluated on a $5 \times 5 \times 1$ q-point grid. The resulting phonon frequencies and eigenvectors were used to compute the one-phonon thermal diffuse scattering intensity using the phonopy package [45].

**Acknowledgements**

We thank C. W. Hicks, A. P. Mackenzie, Ž. Skoko and A. Ramires for helpful comments and discussions.

**Funding statement**

The work in Zagreb was supported by the Croatian Science Foundation through Grant No. UIP-2020-02-9494, and the Croatian Ministry of Science, Education and Youth. The work at the University of Minnesota was funded by the US Department of Energy through the University of Minnesota Center for Quantum Materials, under Grant No. DE-SC-0016371. We acknowledge Agastya HPC cluster at IIT Jammu for providing computational resources. Work at Argonne (RO, SR) was supported by the DOE Office of Science, Basic Energy Sciences, Materials Sciences and Engineering Division. This research used resources at the Advanced Photon Source, a DOE Office of Science user facility operated for the DOE Office of Science by Argonne National Laboratory under Contract No. DE-AC02-06CH11357. A portion of this research used resources at the Spallation Neutron Source (SNS), a DOE Office of Science user facility operated by the Oak Ridge National Laboratory (ORNL). The beamtime was allocated to CORELLI on proposal numbers IPTS-27428 and IPTS-33230.

**Author Contributions**

DP and MG designed the study. NK and DS prepared samples. MS, RS, IB, BKO, XH, AM, FY, MJK, SR, RO and DP performed neutron and x-ray diffuse scattering measurements and analysed data. IB, BKO, MS and ET performed high-resolution x-ray diffraction measurements and analysed data. AP, MS and TB performed first-principles calculations. DP, MG, AP, IB and MS wrote the manuscript, with input from all coauthors.

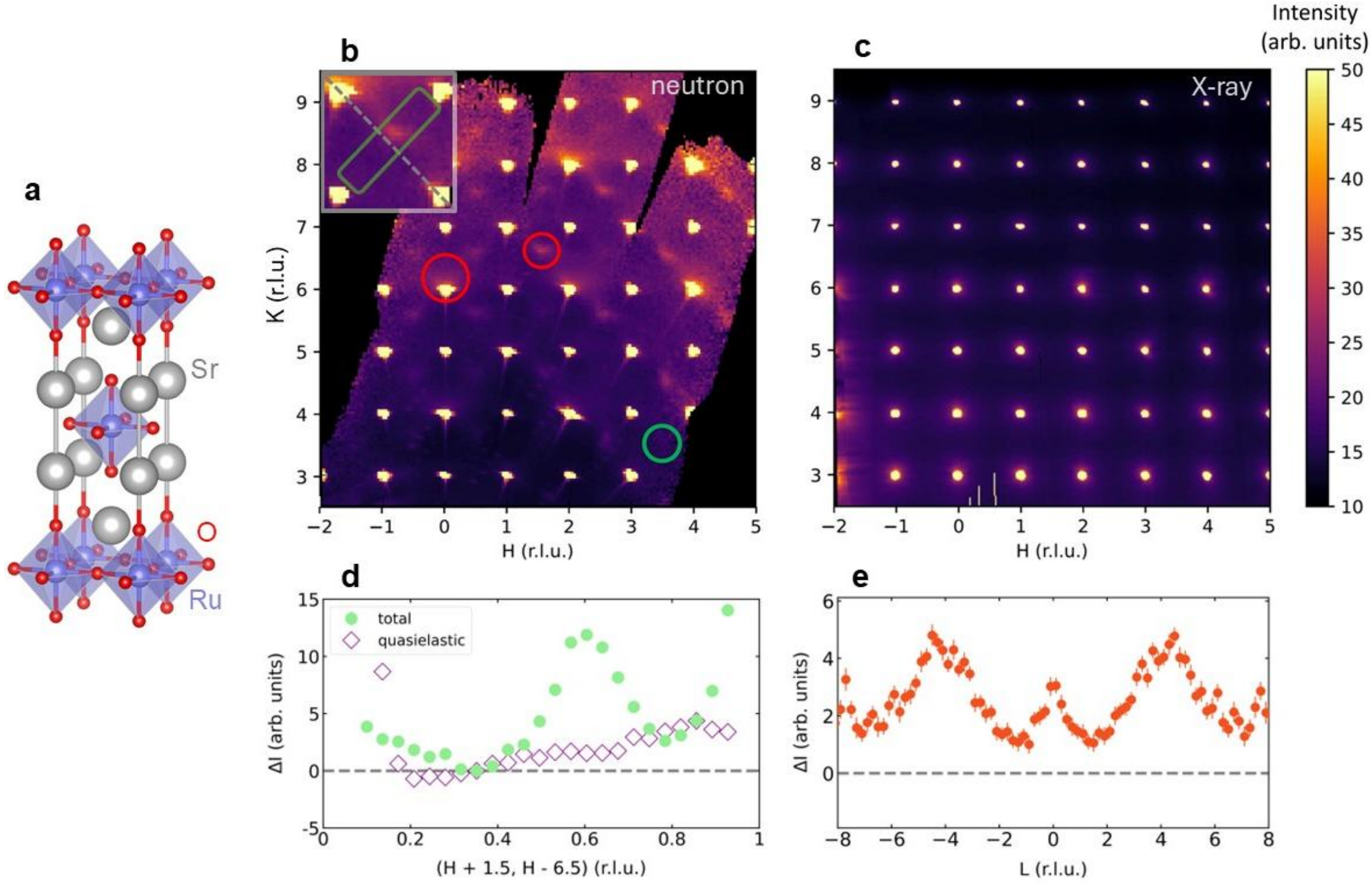

**Fig. 1 | Neutron and x-ray diffuse scattering in strontium ruthenate.** (**a**) The highly symmetric (average) body-centered tetragonal structure of $Sr_2RuO_4$. (**b**) Neutron total scattering data in the *HK*-plane at 5 K, integrated along [001] from $L = -12$ to 12 reciprocal lattice units (r.l.u.). Red circles: Diffuse features are around some half-integer positions and integer (Bragg) positions on the high-wavevector side. Green circle: an extinction, i.e., a half-integer position where no scattering is observed, which is key for the reconstruction of the atomic displacement pattern. Inset: region around the (1.5 6.5) position, showcasing the asymmetry of the diffuse feature. The green square marks the integration range for (d). (**c**) Diffuse x-ray scattering data, integrated in the same *L*-range as in (b), measured at 30 K. No diffuse features are observed within sensitivity, which implies that the short-range correlations predominantly involve oxygen displacements. (**d**) Comparison of background-subtracted quasi-elastic and energy-integrated (total) neutron scattering, demonstrating the near-absence of static correlations; see Extended Data Fig. 3 for a full reciprocal space map of the quasi-elastic scattering. (**e**) Background-subtracted *L*-dependence of the neutron scattering intensity around $(H, K) = (1.5, 6.5)$ shows a smooth modulation, but no sharp peaks, consistent with mostly two-dimensional correlations. The background estimate was obtained from an integration box with similar size, but at the slightly different position (1.5, 6); see Extended Data Fig. 1 for raw data.

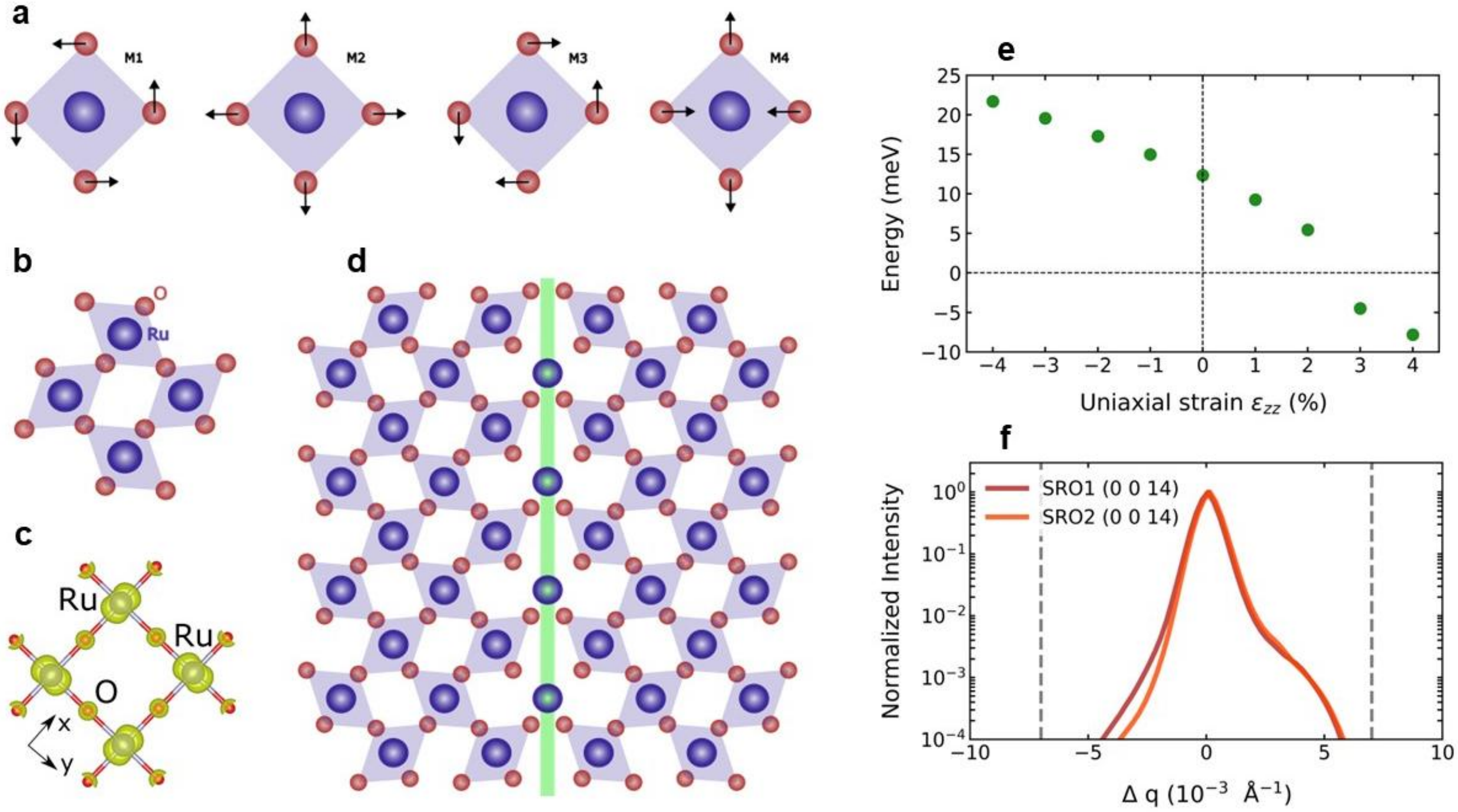

**Fig. 2 | Atomic displacement pattern and absence of extrinsic disorder. (a)** The four irreducible in-plane distortions of the Ru-O plaquette. **(b)** Shear-like distortion that consists of equal weights of M1 and M4 and is consistent with the neutron scattering data (see Methods and Extended Data Fig. 2 for details on refinement). **(c)** orbital order from *ab initio* calculations, where the isosurface represents partial charge density associated with states at the zone-centre valence band maximum (see Methods for details). The charge density distribution underlines alternating short and long Ru-O bonds in close correspondence with the observed distortion. **(d)** Defect structure that includes the monoclinic-type distortion as well as local oxygen sublattice contraction, which decays away from the central line. The contraction explains the observed reciprocal space asymmetry of the diffuse features. (**e**) Softening and instability of the octahedral rotation phonon mode with tensile strain along the crystallographic c-axis, obtained from *ab initio* calculations. (**f**) Quantification of internal strains along the c-axis in two SRO crystals using high-resolution x-ray diffraction at the (0 0 14) Bragg peak. The data are deconvoluted from a Lorentzian resolution function (see Methods and Extended Data Fig. 3 for raw data). The volume fraction of SRO that is deformed more than ±0.1% (vertical lines) is clearly well below 1%, implying that the rotational mode softening cannot be responsible for the diffuse scattering signal.

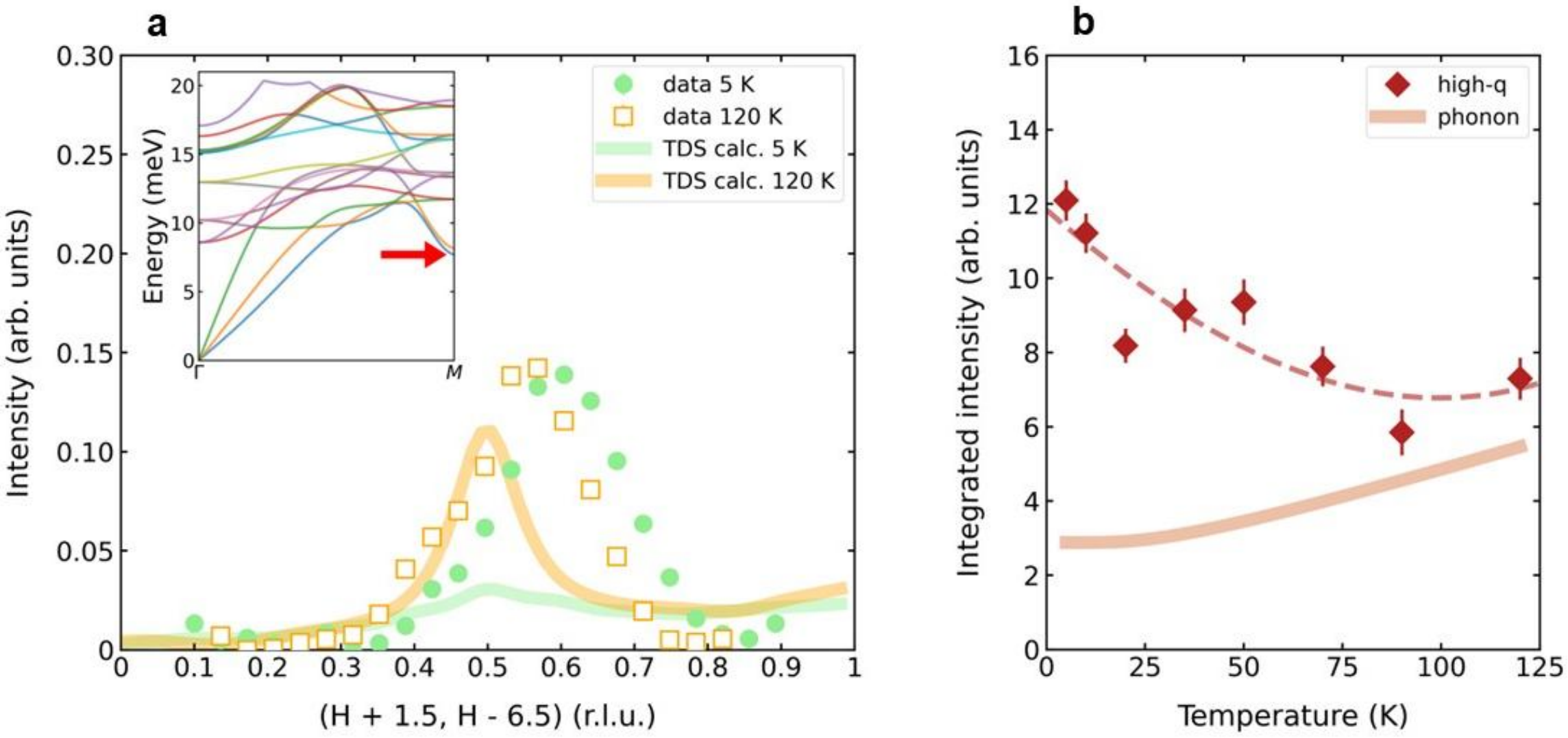

**Fig. 3 | Temperature dependence of diffuse scattering.** (**a**) Comparison of neutron total scattering data to a calculation of phonon scattering (TDS) at two temperatures, demonstrating that the observed asymmetric features are inconsistent with scattering due to phonons, especially at low temperatures. Inset: low-energy (<20 meV) *ab initio* phonon dispersions used in the TDS calculation, which are in good agreement with experiment [17,24] and assumed to be temperature-independent. The octahedral rotation mode at the zone boundary $M$ point (arrow) results in the diffuse scattering peak, and the TDS decrease upon cooling originates from the Bose occupation factor. (**b**) Temperature dependence of the high-wavenumber component (above $q$ = 0.5 r.l.u.) of the diffuse scattering around (1.5, 6.5) (dashed line is a guide to the eye). Line: the contribution from the softest octahedral rotation phonon mode at 8 meV; the energy of this mode is known to be nearly temperature-independent below 100 K [17,24].

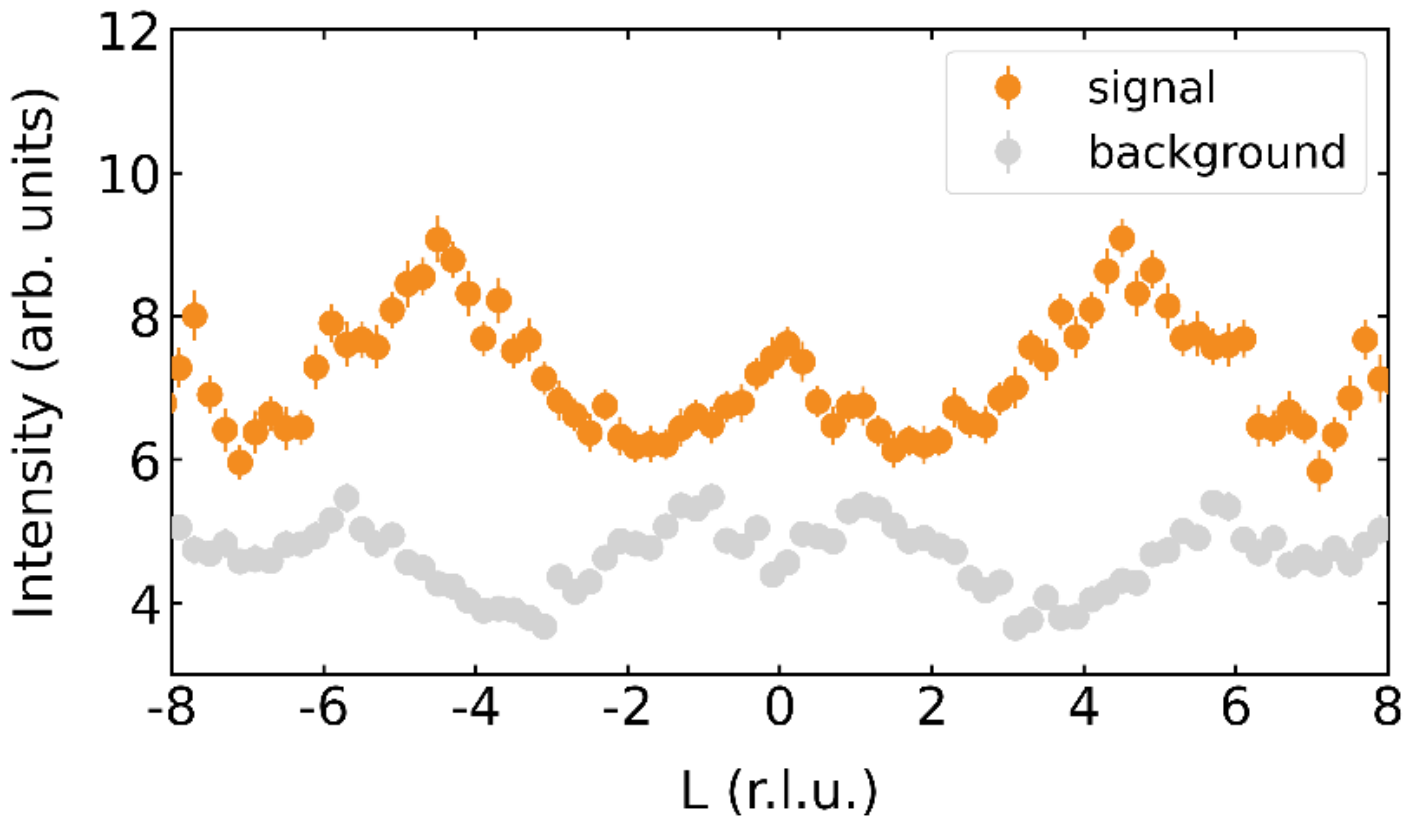

Extended Data Fig. 1 | Raw data for $L$ dependence of diffuse scattering at the (1.5, 6.5) position, at 5 K. The background (grey) is obtained from a neighbouring position with no distinct features, and sharp peaks at higher $L$ are likely aluminium powder diffraction rings from the sample environment.

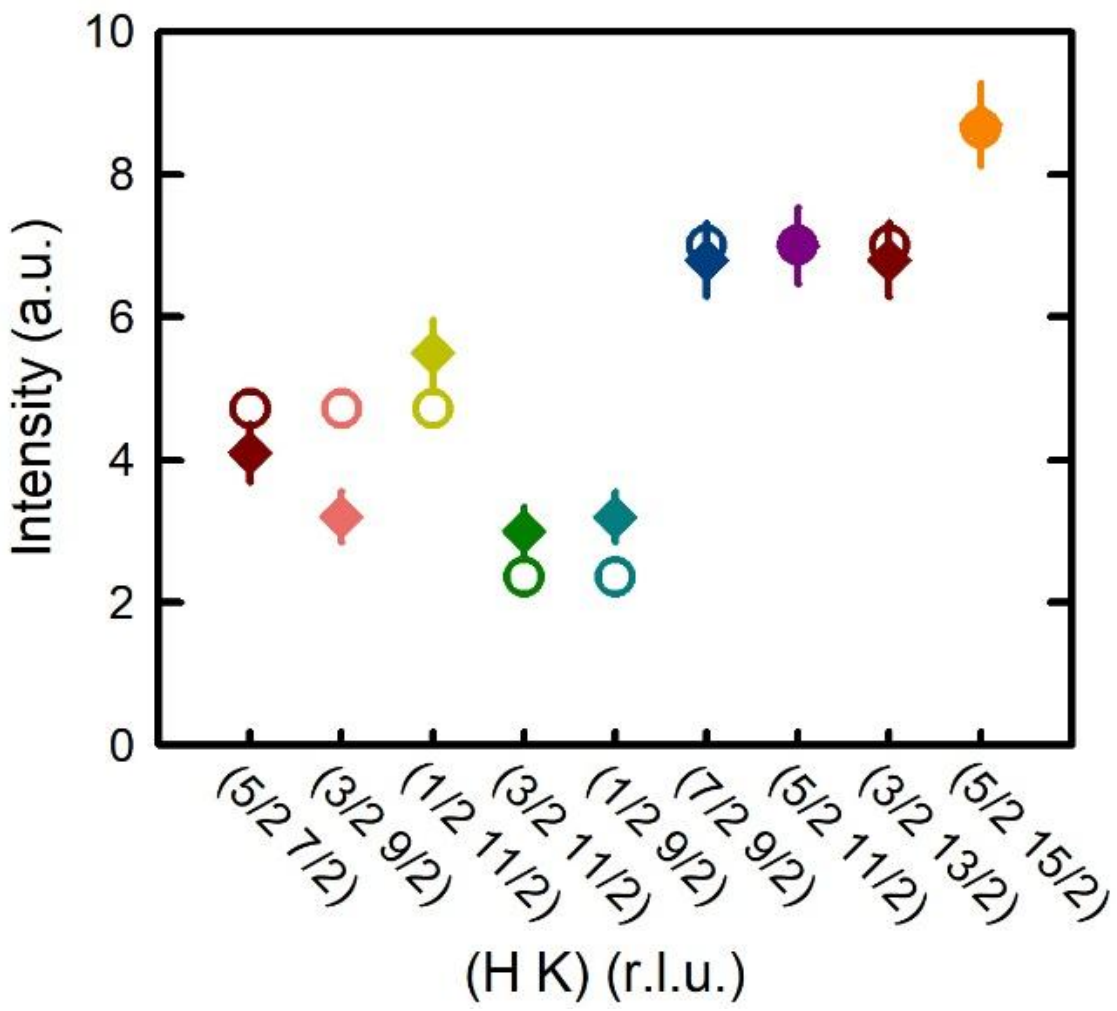

Extended Data Fig. 2 | Refinement of diffuse scattering intensities at nine different positions (full symbols, error bars represent 1 s.d.) to a two-dimensional distortion model that includes M1 and M4 irreducible distortions (empty symbols, see Methods for model an fit details). The best fit is obtained for equal weights of M1 and M4, which implies a shear-like distortion (Fig. 2b).

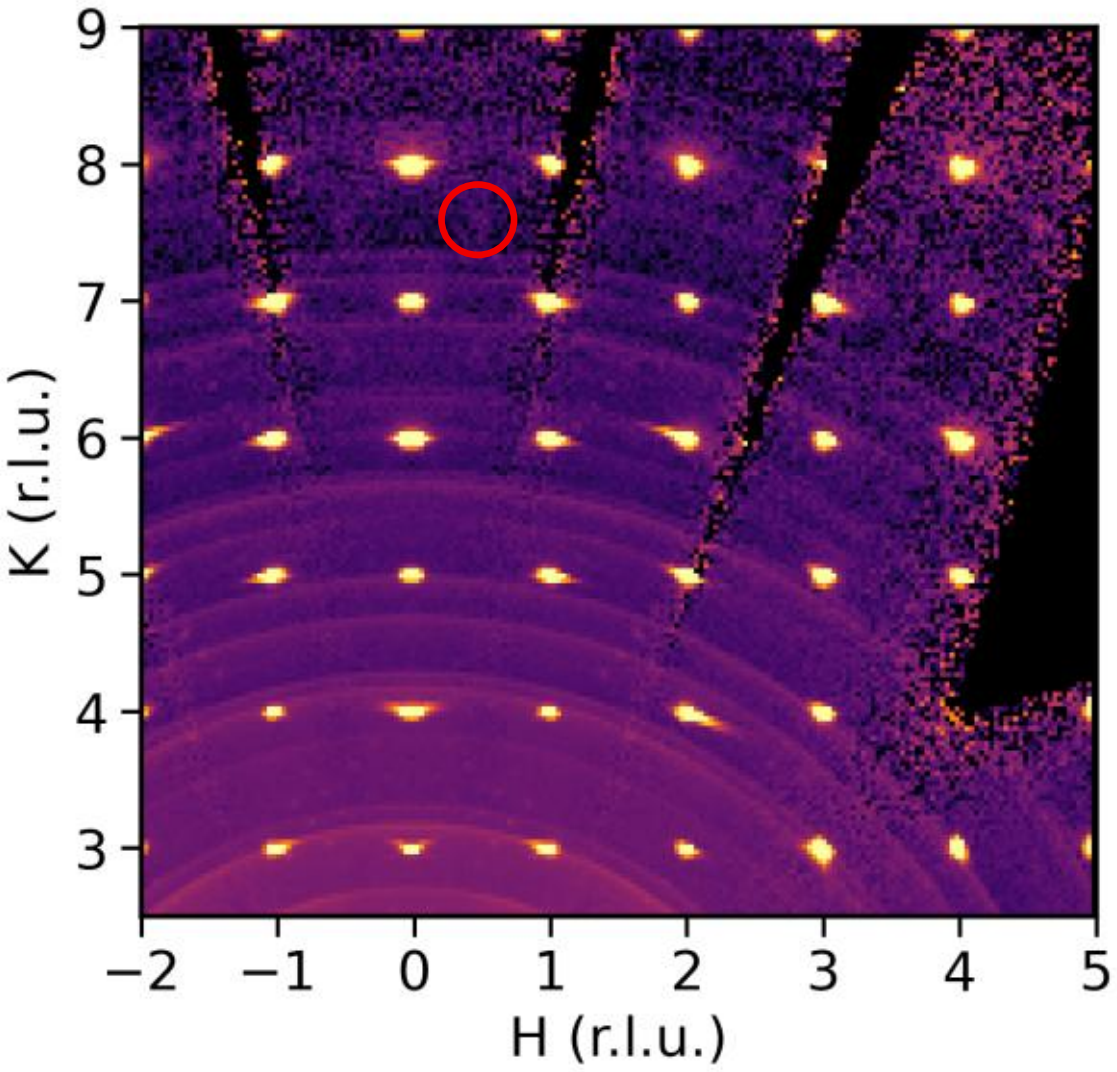

Extended Data Fig. 3 | Quasi-elastic diffuse scattering channel at 5 K, shown with the same binning as the total (energy-integrated) channel in Fig. 1b. Half-integer features are barely visible at high wavevectors (red circle), where the energy resolution is around 2 meV.

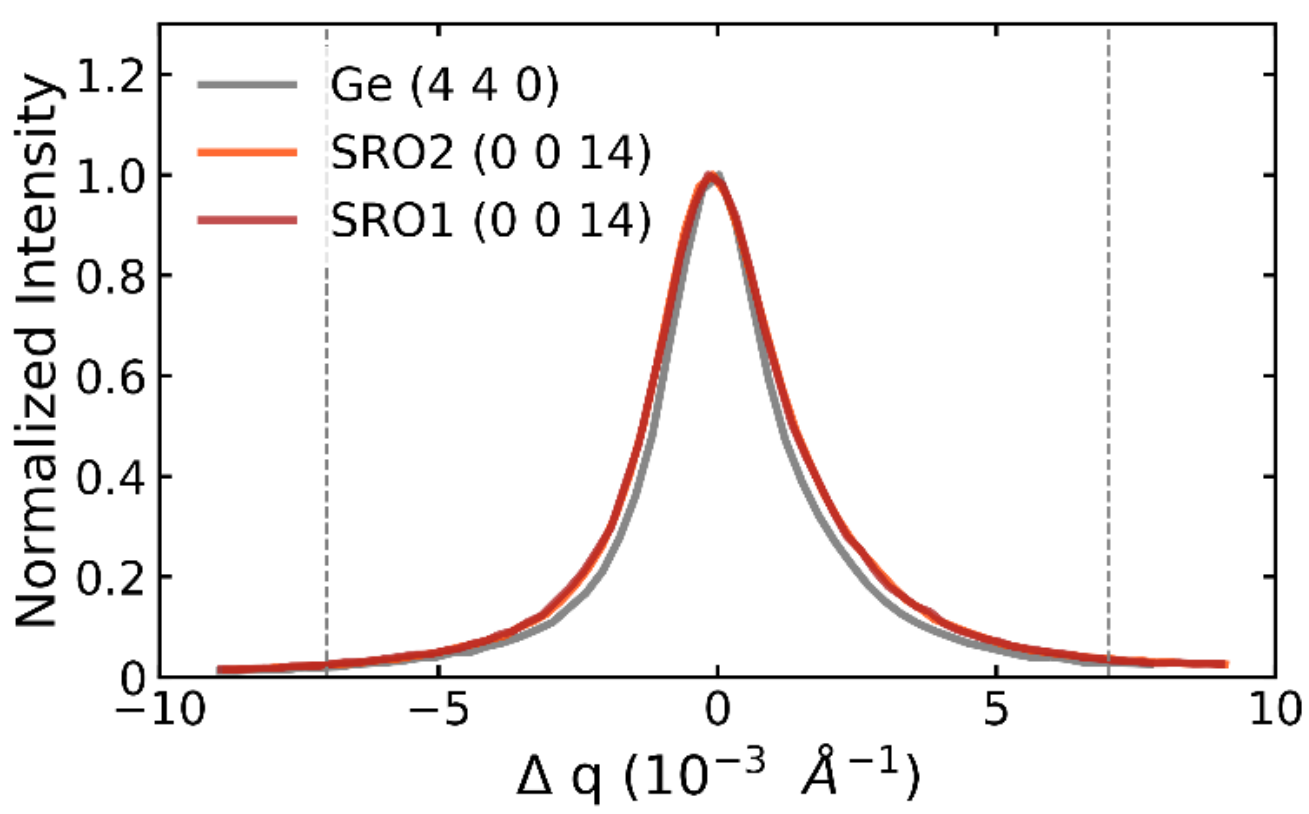

Extended Data Fig. 4 | Raw high-resolution x-ray diffraction data for two SRO crystals (red and orange, overlapping lines), compared to a commercial germanium single crystal (grey full line). The vertical dashed lines correspond to lattice deformation levels of 0.1%.